\documentclass[a4paper]{article}

\usepackage{latexsym}
\usepackage{amsmath,amssymb,theorem}
\usepackage[english]{babel}

\newtheorem{teo}{Theorem}[section]

\newtheorem{lema}[teo]{Lemma}
\newtheorem{coro}[teo]{Corollary}

{\theorembodyfont{\rmfamily }
\newtheorem{defi}[teo]{Definition}

\newtheorem{remark}[teo]{Remark}}

\def\demo{\noindent \textit{Proof: }}
\def\d{\mathrm{d} }

\begin{document}

\title{Lovelock's theorem revisited}

\author{Alberto Navarro, Jos\'{e} Navarro} 
\footnotetext[1]{Department of Mathematics, University of Extremadura, Avenida de Elvas s/n, 06071, \\
Badajoz, Spain. \\
\textit{Email addresses:} navarrogarmendia@unex.es\\ }

\maketitle

\begin{abstract}
Let $(X, g)$ be an arbitrary pseudo-riemannian manifold. A celebrated result by Lovelock (\cite{Primero}, \cite{Lovelock}, \cite{LovelockDimension})
gives an explicit description of all second-order natural (0,2)-tensors on $X$, that satisfy the conditions of being symmetric and divergence-free. Apart from the dual metric, the Einstein tensor of $g$ is the simplest example.

In this paper, we give a short and self-contained proof of this theorem, simplifying the existing one by formalizing the notion of derivative of a natural tensor.
\end{abstract}

\section*{Introduction}\label{intro}

Let $(X,g)$ be a pseudo-riemannian manifold of dimension $n$. A classical problem, mainly motivated
 by the aim of determining the field equation for a gravitational theory, is to compute
divergence-free (0,2)-tensors that can be intrinsically obtained from $g$. The example par excellence is that of the Einstein
tensor, that fits into the field equation of General Relativity.

The notion of tensor ``intrinsically'' constructed from a metric corresponds to
that of \textit{natural tensor},
and methods from the theory of natural operations (\cite{Kolar}) sometimes allow to compute,
more or less explicitly, all possible natural tensors of a certain kind.

A first example of so is a classical result saying that, apart from the Einstein tensor,
there is no other symmetric and divergence-free natural (0,2)-tensor with the condition of being linear on the
second derivatives of the metric (\cite{Autocita}).

Later on, Lanczos (\cite{Lanczos}) found a more complicated divergence-free natural (0,2)-tensor.
This \textit{Lanczos tensor} is  quadratic
on the second derivatives of the metric, but only exists if $X$ is of dimension strictly greater than 4.

No other divergence-free natural (0,2)-tensor was known until the beginning of the 70's, when D. Lovelock proved in a series of papers (\cite{Primero},
\cite{Lovelock}, \cite{LovelockDimension}) his well-known result: he listed all the symmetric and divergence-free natural (0,2)-tensors with coefficients depending on
second derivatives of the metric.

However, in spite of its simple and clear statement, the proof of this result is rather long and intricate. In this paper, we apply the theory of natural tensors (\cite{Epstein}, \cite{Kolar}, \cite{Stredder}) to obtain a short proof, that simplifies the original arguments.

To do so, we introduce the notion of derivative of a second-order natural tensor (Section \ref{sec:2}), that allows us to easily prove a general statement (Theorem \ref{Homogenous}). Lovelock's result (Theorem \ref{Teorema}) readily follows as a corollary, by simply doing some standard computations. Here, the main advantage is the use of the invariant theory for the orthogonal group, that allows to avoid the cumbersome ``integration" argument that
appears in the original papers.

For the sake of completeness, we begin by an elementary exposition of the
topics needed from the theory of natural operations, equivalent to the standard categorical approach (\cite{Kolar}).

\section{Second-order natural tensors}
\label{sec:1}

Let $X$ be a smooth manifold of dimension $n$, $M \to X$ be the bundle of pseudo-riemannian metrics (with a prescribed signature), $J^2M \to X$ be its bundle of 2-jets, and $T_p^q \to X$ be the bundle of $(p,q)$-tensors on $X$.

These bundles $M$ and $T_p^q$ are \textit{natural}, in the sense that local diffeomorphisms of $X$ act on them.

\begin{defi} A second-order\footnote{All natural tensors considered in this paper will be second-order.} \textbf{$(p,q)$-natural tensor} on $X$
is a morphism of natural bundles $A \colon J ^2 M \to T_ p^q$; that is, a morphism of bundles commuting with the action of local diffeomorphisms (see details on \cite{Epstein}, \cite{Kolar} or \cite{Autocita}).
\end{defi}

\medskip Loosely speaking, such a natural tensor $A$ is a construction that assigns, to any pseudo-riemannian metric $g$
on an open set $U \subset X$, a $(p,q)$-tensor $A(g)$ on $U$, satisfying the following conditions:
\begin{enumerate}
 \item \textit{Locality:} If $V \subset U$ is an open set, then $A(g)_{|V}=A(g_{|V})$.

 \item \textit{Naturalness:} For any diffeomorphism $\tau \colon U \to V$ between
open sets of $X$, it holds:
\begin{equation*}\label{Naturalness}
            A( \tau^* g) = \tau^* (A(g))
\end{equation*}

 \item \textit{Second-order:} At any point $x \in X$, the value of the tensor $A(g)$ only
depends on the metric $g$ and its first and second derivatives at $x$.

That is to say, there exists ``universal'' smooth functions $A^{i_1 \ldots i_q}_{j_1 \ldots j_p}$ such that,
in any chart $x^1 , \ldots , x^n$:
$$ A(g) = \sum_{i_1,  \ldots , j_p} A^{i_1 \ldots i_q}_{j_1 \ldots j_p} (g_{ab}  , g_{ab,c} , g_{ab,cd} ) \,
\partial_{ x_{i_1}} \otimes \ldots \otimes \mathrm{d} x^{j_p} $$ where $g_{ab,c} := \partial g_{ab} / \partial x^c $,
$g_{ab,cd} := \partial^2 g_{ab} / \partial x^c \partial x^d$.
\end{enumerate}

\medskip If $x \in X$ is a point,
there exists local diffeomorphisms mapping $x$ to any other point on $X$. Then, from the naturalness condition it follows that a natural tensor is determined by its value on the Taylor expansions of metrics at $x$.

In normal coordinates, the second-order Taylor expansion of a metric $g$  at $x$ is $\delta_{ab} + g_{ab,cd}$, where $ \delta$ is the
diagonal matrix with as many 1 and $-1$ as the signature.

These remarks say that giving a natural tensor is the same as giving the following
collection of smooth functions:
\begin{equation}\label{Bijection} f^{i_1 \ldots i_q}_{j_1 \ldots j_p} (g_{ab,cd}) := A^{i_1 \ldots i_q}_{j_1 \ldots j_p} ( \delta_{ab} , 0 , g_{ab,cd})
\end{equation}

Normal coordinates are well-defined up to a transformation of the orthogonal group $O$
of $\delta$. Therefore, these functions $f^{i_1 \ldots i_q}_{j_1 \ldots j_p}$ still satisfy an equivariant condition. To state it precisely, we introduce normal tensors at $x$:

\begin{defi} The space of \textbf{normal tensors} at a point $x\in X$ is the vector subspace $ N \subseteq \otimes^4 T^*_x X$
whose elements $T$ have the following symmetries:
\begin{enumerate}
\item they are symmetric in the first two and last two indices:
$$ T_{ab cd} = T_{ba cd} \quad , \quad T_{ab cd} = T_{ab dc} $$
\item the cyclic sum over the last $3$ indices is zero:
$$ 0 = \sum_{(b c d )} T_{abcd} : = T_{ab cd} + T_{a d b c } + T_{a c d b} \ .$$\end{enumerate}
\end{defi}

Any metric $g$ on a a neighbourhood of a point $x \in X$ defines a normal tensor at $x$: if $z^1 , \ldots , z^n$
 are normal coordinates for $g$ at $x$, then:
$$ \sum_{a,b,c,d } g_{ab,cd}(x) \  \d z^a \otimes \d z^b \otimes \d z^c \otimes \d z^d $$
is a well-defined normal tensor at $x$.

These considerations prove the following theorem:

\begin{teo}[\cite{Stredder}]\label{Basico} The map $A \mapsto f$ establishes an isomorphism of $\mathbb{R}$-vector
spaces:
$$\{\begin{matrix} \mbox{Natural $(p,q)$-tensors}  \end{matrix}
\} \simeq
\left[ \begin{matrix} \mbox{ Smooth $O$-equivariant maps } \\
f \colon N \to T_{p,x}^q \end{matrix} \right]$$
where $T_{p,x}^q $ is the vector space of $(p,q)$-tensors at $x \in X$ and $O$ is the orthogonal group for the diagonal metric $\delta$
at $x$.

A natural tensor $A $ is said to be \textbf{polynomial} (resp. \textbf{homogenous of degree }$k$) if its associated
smooth map is polynomial (resp. homogenous of degree $k$).
\end{teo}

\subsection{Derivative of a natural tensor}
\label{sec:2}

Let $N$ and $E$ be two $\mathbb{R}$-vector spaces of finite dimension (they will later be the vector space of normal tensors and a tensor power of $T_xX$).

If $f \colon N \to E $ is a smooth map, then at each point $p \in N$ we have the tangent linear map:
$$ N = T_p N \xrightarrow{ d_pf } T_{f(p)} E = E $$ and therefore,
we can consider the smooth map:
\begin{align*} N \xrightarrow{df} \mathrm{Hom}_{\mathbb{R}-lin} ( N , E) = E \otimes N^*  \quad , \quad p \mapsto d_pf
\end{align*}

Iterating the process, we obtain the higher derivatives:
\begin{align*} N &\xrightarrow{d^mf}  E \otimes N^* \otimes \stackrel{m}{\ldots} \otimes N^*
\end{align*}

Choosing a basis, if $f^i(y_1 , \ldots , y_r)$ are the components of  $f$, then:
\begin{equation}\label{Expresion} d_pf \equiv \left( \frac{\partial f^i}{\partial y_a} (p) \right) \quad , \quad
df \equiv \left( \frac{\partial f^i}{\partial y_a} \right) \quad , \quad d^m f \equiv \left( \frac{\partial^m f^i}{\partial y_{a_1} \ldots \partial y_{a_m}} \right) \end{equation}

Therefore, as partial derivatives commute, these higher derivatives take values on the symmetric powers:
\begin{align*} N &\xrightarrow{d^mf} E \otimes S^m N^*
\end{align*}

\begin{remark}\label{DerivadaIterada} It can be checked that, if $N$ and $E$ are linear representations of the orthogonal group $O$ and
 $f \colon N \to E$ is $O$-equivariant, then $df $ is also $O$-equivariant.

In case $f$ is a homogenous polynomial of degree $k$, then $d^m f$ is a homogenous polynomial of degree $k-m$.

Therefore, if $m = k$,
 the map $f \mapsto d^k f$ establishes an isomorphism of $\mathbb{R}$-vector spaces:
$$ \mathrm{Hom}_O ( S^k N , E )  =  \mathrm{Hom}_O (\mathbb{R} , E \otimes S^k N^* )$$
where $\mathrm{Hom}_O$ is the vector space of $\mathbb{R}$-linear $O$-equivariant maps.
\end{remark}

\bigskip
Let $S^2$ be the vector space of symmetric $(0,2)$-tensors at a point  $x \in X$ and
recall $N \subset \otimes^4 T_x^* X$ stands for the vector space of normal tensors at $x$.

By Theorem \ref{Basico}, a symmetric (0,2)-natural tensor $A$ is defined by a smooth $O$-equivariant map:
$$ f \colon N \to S^2 $$

\begin{defi} The \textbf{derivative} of a symmetric (0,2)-natural tensor $A$ is the (0,6)-natural tensor
$A' $ defined by the smooth map:
$$ df \colon N \to S^2 \otimes N^* $$

Analogously, the \textbf{higher derivatives} $ A^{m)} $ 
are $(0, 4m +2)$-natural tensors defined by the smooth maps:
$$ d^m f \colon N \to S^2 \otimes S^m N^*   $$
\end{defi}

\begin{remark}\label{CoordenadasDerivada} In a local chart, if $A^{ij}$ are the components of a natural tensor $A$, then the components
$A^{ij; ab,cd} $ of the natural tensor $A'$ are precisely:
\begin{equation}\label{ExpresionDerivada}
 A^{ij; ab , cd} = \frac{\partial A^{ij}}{\partial g_{ab,cd}}
\end{equation}
as follows from the general
expression (\ref{Expresion}) of $df$ and because:
$$ \frac{\partial (A^{ij} (\delta_{ab}, 0 , \, \_ \, ))}{\partial g_{ab,cd}} =  \frac{\partial A^{ij}}{\partial g_{ab,cd}} \, (\delta_{ab} , 0 , \, \_ \, ) $$ \end{remark}

\begin{teo}\label{BasicoHomog}
The map $A \mapsto d^k f$ establishes an isomorphism of $\mathbb{R}$-vector spaces:
\begin{align*}
 \left[ \begin{matrix} \mbox{Symmetric, natural (0,2)-tensors} \\ \mbox{ homogenous of degree  $k$} \end{matrix}
\right]  \simeq (S^2 \otimes S^k N^* )^O
\end{align*} where $  (S^2 \otimes S^k N^* )^O$ is the subspace of vectors invariant under the action of the orthogonal group $O$.
\end{teo}

\demo By Theorem \ref{Basico}, the map $A\mapsto f$ is an isomorphism:
$$ \left[ \begin{matrix} \mbox{Symmetric, natural (0,2)-tensors} \\ \mbox{ homogenous of degree  $k$} \end{matrix}
\right]
\simeq \mathrm{Hom}_O ( S^k N , S^2  ) $$
and, by Remark \ref{DerivadaIterada}, the map $f \mapsto d^k f$ establish:
$$ \mathrm{Hom}_O ( S^k N , S^2  ) = \mathrm{Hom}_O ( \mathbb{R} , S^2 \otimes S^k N^* ) = (S^2
\otimes S^k N^* )^O $$ \hfill $\square$

\subsection{Divergence-free tensors}

\begin{defi} A natural $(0,2)$-tensor $A$ is said to be \textbf{divergence-free}  if, for any metric $g$, it holds: $$C^2 _1 (\nabla_g A(g))=0 $$ where $\nabla_g$ is the Levi-Civita connection of $g$ and $C_1^2$ denotes contraction of the first covariant and second contravariant indices.

In a local chart, it amounts to saying that, for any metric $g$, the functions
$A^{ij} = A^{ij}(g_{ab}, g_{ab,c}, g_{ab,cd})$ satisfy the equation:
$$ \nabla_j A^{ij} = 0$$ where $\nabla$ is the Levi-Civita connection of $g$.
\end{defi}

\begin{lema}\label{LovelockLemma} If a natural $(0,2)$-tensor $A$ is divergence-free, then its derivative $A'$ is a natural tensor
satisfying the following linear symmetry:
\begin{equation}\label{LovelockId}
0 = \sum_{( j \, c \, d)} A^{ij; ab,cd} := A^{ij;ab,cd} + A^{id;ab,jc} + A^{ic; ab, dj}
\end{equation}

\end{lema}

\demo In a local chart:
\begin{align*}
 \nabla_j A^{ij} &= \frac{\partial A^{ij}}{\partial g_{ab,cd}}\, g_{ab.cdj} +  F(g_{ab} , g_{ab,c}, g_{ab,cd} ) \\
 &= \frac{1}{3} \left( \frac{\partial A^{ij}}{\partial g_{ab,cd}} +
 \frac{\partial A^{id}}{\partial g_{ab,jc}} + \frac{\partial A^{ic}}{\partial g_{ab,dj}} \right) g_{ab,cdj} + F(g_{ab} ,
 g_{ab,c} , g_{ab,cd})
\end{align*}where we use summation over repeated indices.

Therefore, the condition $\nabla_j A^{ij} = 0$, valid for any metric, implies:
$$ \frac{\partial A^{ij}}{\partial g_{ab,cd}} +
 \frac{\partial A^{id}}{\partial g_{ab,jc}} + \frac{\partial A^{ic}}{\partial g_{ab,dj}}  = 0$$
that, because of (\ref{CoordenadasDerivada}), is equivalent to the thesis.\hfill $\square$

\begin{defi} Let $\mathrm{Div}^m \subset S^2 \otimes S^m N^*  $ be the vector subspace whose elements satisfy the following symmetry:
\begin{align}\label{Simetria}
 0 &= \sum_{(j\, a_3 a_4 )} T^{ij a_1  \ldots a_{4m}}
\end{align}
\end{defi}

\begin{teo} \label{Homogenous} The map $A \mapsto d^kf$ defines an inclusion:\footnote{Although we will not use it here, this map is
indeed an isomorphism: if $A'$ satisfies symmetry (\ref{LovelockId}), then $A$ is divergence-free (see \cite{Primero}).}
 $$ \left[ \begin{matrix} \mbox{Symmetric, natural (0,2)-tensors, } \\ \mbox{ homogenous of degree $k$ and
divergence-free } \end{matrix} \right]  \subseteq ( \mathrm{Div}^k )^{O} $$
\end{teo}

\demo It follows from Theorem \ref{BasicoHomog} and Lemma \ref{LovelockLemma}. \hfill $\square$

\section{Lovelock's Theorem}

The rest of the paper is devoted to compute a basis for the $\mathbb{R}$-vector space $( \mathrm{Div}^k )^{O}$.

\begin{lema}\label{Symmetries} $\mathrm{Div}^m$ is the vector subspace of $\otimes^{4m+2} T_xX$ whose elements are
tensors with the following symmetries:
\begin{enumerate}
\item They are symmetric in each pair of indices $a_{2h-1}  a_{2h} $ for $h = 1 , \ldots , 2m+1 $.

\item They are symmetric under the interchange of  the pair  $a_{2h-1} a_{2h} $ with the pair $a_{2l-1} , a_{2l} $, for $h,l = 1 , \ldots 2m+1$.

\item The cyclic sum of any three consecutive indices is zero.
\end{enumerate}
\end{lema}

\demo Symmetry number \textit{1} is clear from the definitions and symmetry number \textit{3} easily follows from the other two.

To check symmetry number \textit{2}, let us first prove that,
if $T \in N$ is a normal tensor at $x \in X $, then $T_{ijkl} = T_{klij}$:
\begin{align*}
T_{ijkl} &= - T_{iljk } - T_{iklj \ldots} = -T_{lijk } - T_{kijl } \\
&= T_{lkij } + T_{ljki } + T_{klij } + T_{kjli } \\
&= 2 T_{klij } + T_{jlki } + T_{kjli } \\
&= 2 T_{klij } - T_{jilk } - T_{jkil } + T_{kjli } = 2T_{klij } - T_{ijkl }
\end{align*}

Now, an analogous reasoning proves symmetry number \textit{2} in full generality. \hfill $\square$

\begin{lema}\label{DivGrande}
Recall $X$ has dimension $n$. It holds:
$$ \dim_\mathbb{R}(\mathrm{Div}^m )^O =  \begin{cases}
 1 &\mbox{ if } m \leq  \left[ \frac{n-1}{2} \right]  \\
  0 & \mbox{ in other case. }  \\
\end{cases}$$
\end{lema}

\demo If $m$ is greater than the integer part of $(n-1)/2$, elements of $\mathrm{Div}^m$
are tensors with $4m +2 >   2n$ indices, so each
component of any such a tensor has, at least, three repeated indices.

Due to symmetries \textit{1} and \textit{2} of Lemma \ref{Symmetries},
any configuration of indices can be reduced to one of the following:
$$ a a a b c \ldots \qquad , \qquad a b a c a d \ldots $$

Using symmetries again, it is easy to check that both configurations are proportional.
The first one is clearly zero, due to symmetry \textit{3}, so we conclude $\mathrm{Div}^m=0$.

\medskip Now, let $m$ be lesser or equal than the integral part of $(n-1)/2$.
By the Main Theorem of the invariant theory for the orthogonal group,
total contraction of indices are a system of generators for $(\mathrm{Div}^m )^{O}$.

Assume $m= 1$, the general case being analogous; we are going to prove that any total contraction of indices is proportional to:
$$ (1,2) (3,4) (5,6) $$ where $( \, , \, )$ means contraction of the indices inside.

Given a total contraction, suppose  the index 1 is not contracted with the 2. It has to be contracted with one of the
others: $ 3,4,5,6 $.  Due to symmetries \textit{2} and \textit{3}, we can assume, with no loss of generality, that it is the index 3.

Now, if index 2 is contracted with index 4, then symmetry \textit{3} shows that the pair of contractions (1,3)(2,4) is
proportional to the pair (1,2)(3,4), and we are done.

In other case, we can assume  index 2 is contracted with index 5. By the previous argument, (2,5)(4,6)
is proportional to (2,4)(5,6). Therefore, $(1,3)(2,5)(4,6)$ is proportional to $(1,3)(2,4)(5,6)$ and so to $(1,2)(3,4)(5,6)$. \hfill $\square$

\begin{coro}\label{Polinomialidad}
If a symmetric, natural (0,2)-tensor is divergence-free, then it is polynomial, of degree lesser or equal than $\left[ \frac{n-1}{2} \right]$.
\end{coro}

\demo If $A$ is such a natural tensor, then its $m$-th derivative $A^{m)}$ is a natural tensor defined by the smooth map:
$$ d^m f \colon N \to \mathrm{Div}^m \subset S^2 \otimes S^m N^* $$

By the previous lemma, $d^{m} f=0$ for $m > \left[ \frac{n-1}{2} \right]$, and the statement follows.\hfill $\square$

\begin{lema}\label{FormasValoradas} On a pseudo-riemannian manifold $(X, g)$ of dimension $n$, there exists an isomorphism between the bundle
of $(0,2)$-tensors and the bundle of $(n-1)$-forms with values on $(n-1)$-forms.

Moreover, if $T$ and $\Pi$ are a $(0,2)$-tensor and a form corresponding via this isomorphism, then:
\begin{align}\label{Equivalencia}
T \mbox{ is symmetric } \ \ &\Leftrightarrow \ \ C_1 ^1 (\Pi ) = 0 \\
\label{EquivDiv} \mathrm{div} \, T =0 \ \  &\Leftrightarrow \ \ \d_\nabla \Pi =0
\end{align}
where $\d_\nabla$ is the covariant differential induced by the Levi-Civita connection of $g$ and $C_1^!$ stands for the contraction of the first covariant and first contravariant indices.
\end{lema}

\demo  The isomorphism is locally defined by:
$$ D_1 \otimes D_2 \mapsto i_{D_1} \Omega_g \otimes i_{D_2} \Omega_g $$ where $D_1, D_2$ are vector fields and $\Omega_g$ is any of the two unitary volume forms.

Equivalence (\ref{Equivalencia}) is trivial, so let us check (\ref{EquivDiv}). Let $\{ D_1, \ldots , D_n \}$ be a
local reference of vector fields on a neighbourhood of $x\in X$ such that $(\nabla D_i )_x =0$ for all $i$. If $\{ \theta_0 , \ldots , \theta_n \}$ is the dual basis, we also have $(\nabla \theta_i )_x = 0$, and therefore $\d_x \theta_i = 0$. 

Now, if $\Omega_g = h \d \theta_1 \wedge \ldots \wedge \d \theta_n$, the condition $\nabla \Omega_g = 0$ implies $\d _x h = 0$, so that $\d _x (i_{D_i} \Omega_g ) = 0$, for every $0 \leq i \leq n$.

So, if $ T  = \sum _{i,j} T^{ij}  D_i \otimes D_j$, then:
\begin{align*}
(\nabla T )_x & = \sum _{i,j}  \d_x T^{ij} \otimes (D_i)_x  \otimes (D_j)_x  \quad , \quad (\mathrm{div} T)_x = \left(\sum_j ( \sum _i D_i T^{ij} ) D_j \right)_x \\
\end{align*}
On the other hand, the corresponding form $\Pi  :  = \sum _{i,j} T^{ij} \ i_{D_i} \Omega _g  \otimes i_{D_j} \Omega_g $ satisfies:
\begin{align*}
 (\d_\nabla \Pi)_x & = \sum _{i,j} (\d _x T^{ij}) \wedge (i_{D_i} \Omega _g)_x \otimes (i_{D_{j}} \Omega_g)_x= \\
\phantom{a} &=  (\sum _j (\sum _i D_i f_{ij} )\Omega_g \otimes i_{D_j} \Omega_g )_x= (\Omega _g)_x \otimes
(i_{\mathrm{div} T} \Omega_g)_x
\end{align*} and the statement follows. \hfill $\square$

\medskip As $g$ is non-singular, it establishes an isomorphism:
$$ I_g \colon TX \xrightarrow{\sim} T^*X $$ that can be understood as a natural 1-form with values on 1-forms.

On the other hand, consider the Riemann-Christoffel tensor $R$ of $g$
as a natural 2-form with values on 2-forms:
$$ R \colon TX \wedge TX \to T^* X \wedge T^* X $$

With respect to the wedge product of forms, we can construct the following $(n-1)$-forms with values on $(n-1)$-forms:
\begin{align*}
\widetilde{L}_{k} &:= R \, \wedge \, \stackrel{k}{\ldots} \, \wedge \, R \, \wedge I_g \wedge \ldots \wedge I_g
\end{align*} where $k$ runs from 0 to the integer part of $(n-1)/2$.

Now, it is well known that the conditions of the Levi-Civita connection of $g$ being torsion-free and  the differential Bianchi identity can be restated saying
$ \d_\nabla I_g = 0$ and  $\d _\nabla R = 0 $, respectively. Therefore, it follows, for every $k$:
$$\d_\nabla \widetilde{L}_k = 0 $$

\begin{defi}\label{DefiLovelock} The \textbf{Lovelock's tensors} $L_k$ are the natural (0,2)-tensors that correspond to the
natural forms $\widetilde{L}_k$, via the isomorphism of Proposition \ref{FormasValoradas}.

As examples, it can be checked that $L_0 $ is the dual metric of $g$ and $L_1$ is the
contravariant Einstein tensor of $g$.

In general, each $L_k$ is a homogenous natural tensor of degree $k$.  As the forms $\widetilde{L}_k$ are closed and satisfy $C_1^1( \widetilde{L}_k) = 0$, the tensors $L_k$ are symmetric and divergence-free. \end{defi}

\begin{teo}[Lovelock]\label{Teorema} Let $(X , g)$ be a pseudo-riemannian manifold of dimension $n$ and let $p$ be the integer part of $(n-1)/2$.

The Lovelock tensors:
$$ L_0 , \ldots , L_p $$
are a basis for the $\mathbb{R}$-vector space of second-order
 natural (0,2)-tensors on $X$ that are symmetric and divergence-free.
\end{teo}

\demo If $A$ is such a natural tensor, it is polynomial (Corollary \ref{Polinomialidad}) and we can assume it to be homogenous of degree $k \leq [ (n-1)/2]$.

By Theorem \ref{Homogenous} and Lemma \ref{DivGrande}, $A$ is proportional to the $k$-th Lovelock
tensor $L_k$, and we are done. \hfill $\square$

\label{sec:5}


\begin{thebibliography}{00}

\bibitem{Epstein} Epstein, D.B.A.: \emph{Natural tensors on riemannian manifolds}, J. Diff. Geom., \textbf{10},
 (1975) 631--645.

\bibitem{Kolar} Kol\'{a}r, I., Michor, P.W., Slov\'{a}k, J.: \emph{Natural operations in differential geometry},
(Springer-Verlag, Berlin 1993)

\bibitem{Lanczos} Lanczos, K.: \emph{A remarkable property of the Riemann-Christoffel
tensor in four dimensions}, Ann. of Math. {\bf 39} (1938) 842-850.

\bibitem{Primero} Lovelock, D.: \emph{Divergence-free tensorial concomitants}, Aequat. Math. \textbf{4},
(1970) 127--138.

\bibitem{Lovelock} Lovelock, D.: \emph{The Einstein tensor and its generalizations}, J. Math. Phys. \textbf{12},
(1971) 498--501.

\bibitem{LovelockDimension} Lovelock, D.: \emph{The four dimensionality of space and the Einstein tensor}, J. Math. Phys.
\textbf{13}, (1972) 874--876.



\bibitem{Autocita} Navarro, J., Sancho, J.B. \emph{On the naturalness of Einstein equation}, J. Geom. Phys. \textbf{58},
(2008) 1007-1014.

\bibitem{Stredder} Stredder, P.: \emph{Natural differential operators on riemannian manifolds and
representations of the orthogonal and special orthogonal groups}, J. Diff. Geom. \textbf{10}, (1975) 647--660.



\end{thebibliography}
\end{document}